\documentclass[conference]{IEEEtran-4}

\usepackage{tikz}
\usepackage{amsmath}

\usepackage{subcaption}
\usepackage{booktabs}
\usepackage{multirow}
\usepackage{amssymb}
\usepackage{booktabs}
\usepackage{pifont}
\usepackage{url}
\newcommand{\cmark}{\ding{51}}
\newcommand{\xmark}{\ding{55}}
\usepackage{soul, color, xcolor}

\providecommand{\StepFPM}[3]{%
  \makebox[1.5em][c]{#1}%
  \makebox[0.35em][c]{/}%
  \makebox[1.5em][c]{#2}%
  \makebox[0.35em][c]{/}%
  \makebox[1.5em][c]{#3}%
}
\providecommand{\StepFPMHead}{%
  \makebox[1.5em][c]{F}%
  \makebox[0.35em][c]{/}%
  \makebox[1.5em][c]{P}%
  \makebox[0.35em][c]{/}%
  \makebox[1.5em][c]{M}%
}

\ifCLASSINFOpdf
\else
\fi
\begin{document}
%
\title{An Automated Framework for Extracting Reachable Attack Chains from Cyber Threat Intelligence Reports}

\author{
\IEEEauthorblockN{
Wenbo Hou\textsuperscript{1,2},
Ning Hu\textsuperscript{2,*},
Xueping Wang\textsuperscript{1,3},
Jiahao Gu\textsuperscript{1},
and Wenjian Luo\textsuperscript{1,*}
}

\IEEEauthorblockA{
\textsuperscript{1}Guangdong Provincial Key Laboratory of Novel Security Intelligence Technologies,\\
School of Computer Science and Technology,
Harbin Institute of Technology, Shenzhen, China
}

\IEEEauthorblockA{
\textsuperscript{2}New Network Research Department,
Peng Cheng Laboratory, Shenzhen, China
}

\IEEEauthorblockA{
\textsuperscript{3}Great Bay University, Dongguan, China
}

\IEEEauthorblockA{
24b951083@stu.hit.edu.cn,\quad
hun@pcl.ac.cn,\quad
24b951046@stu.hit.edu.cn,\quad
24b951041@stu.hit.edu.cn,\quad  
luowenjian@hit.edu.cn
}

\IEEEauthorblockA{
\textsuperscript{*}Corresponding authors.
}
}

\maketitle

\begin{abstract}
Cyber Threat Intelligence (CTI) reports richly describe real-world attack processes, but their unstructured narratives cannot be directly used for automated attack-path reasoning.
Existing CTI extraction methods focus on indicators, entities, or TTP labels without modeling the execution conditions and resulting states of each attack step, so the extracted knowledge supports neither state matching nor reachability analysis across multi-stage attack chains.
This paper proposes an automated framework that extracts reachable attack chains by modeling each attack step as an attack unit of preconditions, an attack behavior, and postconditions.
A multi-stage pipeline assisted by large language models (LLMs) extracts attack behavior skeletons, recovers their preconditions and postconditions, normalizes them into predefined predicates, and repairs broken dependencies; the resulting units are compiled into Datalog-style rules for attack-goal reachability reasoning.
On a dataset of 20 CTI reports containing 334 human-validated annotated steps, 
our framework achieves higher annotated-step coverage than representative CTI extraction systems in recovering attack behaviors.
Moreover, by explicitly generating preconditions and postconditions, it produces attack units that are more complete and consistent than those generated by end-to-end LLM baselines.
On the extracted chains, Datalog inference reaches the specified attack goal in 19 of 20 reports, while backward search yields 34 attack paths under the generated rules.
The source code and experimental artifacts are available in an anonymized repository\footnote{\url{https://anonymous.4open.science/r/artifact064b}}.
\end{abstract}


%
\IEEEpeerreviewmaketitle

\section{Introduction}

CTI reports describe real-world attack incidents in detail, covering the full process from initial compromise to final objectives~\cite{CTIBench2024}.
Such reports provide rich knowledge about threat actors, malware behaviors, exploited vulnerabilities, attack tools, and network infrastructure~\cite{TTPDrill2017, SoK2025}.
A central use of this knowledge is downstream attack-path reasoning: given a description of a target system, an analyst wants to know whether an attack goal is reachable, which steps lead to it, and which state dependencies support the corresponding attack paths~\cite{AttacKG2022, AttacKG2025}.
However, most CTI reports are written in unstructured natural language, where attack actions, contextual evidence, and technical details are mixed across paragraphs~\cite{CTIHAL2025}.
Attack steps are often described implicitly or across multiple sentences, making it difficult to identify their temporal order and causal dependencies~\cite{Mining2025, AttackSeqBench2025}.
Although human analysts can mentally reconstruct attack chains from these narratives, automated systems still struggle to transform them into a representation over which such reasoning can actually be performed.

Notably, formal attack-graph reasoning already provides the machinery for this goal.
Logic-based analyzers such as MulVAL~\cite{MulVAL2005} model each attack step as a rule whose body is a set of preconditions and whose head is the resulting security state, and derive reachable attack goals by logical inference.
The bottleneck is therefore not the reasoning engine but the supply of rules: such preconditions and postconditions have traditionally been hand-written by experts, and no existing CTI extraction method produces them automatically from narrative reports.
This rule-supply gap exposes a fundamental limitation of current CTI extraction: existing CTI automation yields descriptive rather than executable knowledge.

\begin{figure*}[t]
  \centering
  \includegraphics[width=\linewidth]{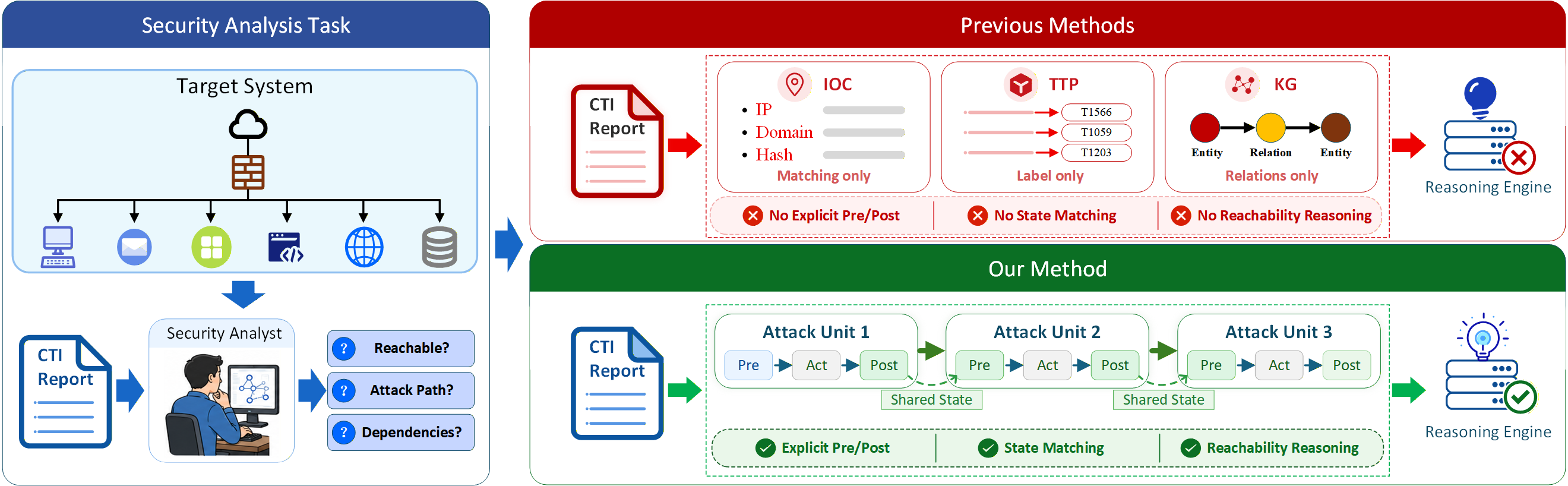}
  \caption{Security analysis task and comparison of previous CTI extraction methods with the proposed approach. Previous methods based on IOCs, TTPs, and knowledge graphs suffer from three key limitations: no explicit preconditions/postconditions, no state matching, and no reachability reasoning. The proposed method represents each attack step as an attack unit with an explicit Pre$\rightarrow$Act$\rightarrow$Post structure, enabling state matching and reasoning about attack-goal reachability.}
  \label{fig:intro_overview}
\end{figure*}

As illustrated in Figure~\ref{fig:intro_overview}, prior methods extract indicators of compromise (IOCs) that can only be matched against observations~\cite{Acing2016}, assign flat MITRE ATT\&CK technique labels without execution conditions~\cite{TTPDrill2017,TTPXHunter2024,mitre2026attackenterprisea}, or build entity-relation and attack knowledge graphs~\cite{AttacKG2022,Extractor2021,ThreatKG2023,CRUcialG2025}, with recent variants using LLMs for graph construction and completion~\cite{LLMTIKG2024,CTINexus2025}.
Graph-based methods add useful structure, such as behavioral sequences, temporal dependencies, and technique associations~\cite{ThreatPilot2024,Automated2025}.
However, they mainly capture the topology among entities and behaviors, rather than the state-transition semantics of each attack step.
They do not specify the conditions under which a behavior becomes executable or the security states produced after execution.
Therefore, their outputs cannot be directly used for state matching, dependency verification, or attack-goal reachability analysis.

These observations expose two gaps that, together, separate current CTI extraction from reasoning-ready knowledge.

\noindent\textbf{Gap 1 (Representation).}
Dominant CTI representations, including IOCs, tactics, techniques, and procedures (TTPs), entity-relation triples, and attack graphs, do not carry explicit preconditions and postconditions.
Without them, an automated reasoner cannot determine whether an attack step is executable under the current system state, nor can it propagate the security states produced by that step to enable subsequent behaviors.
The consequence is concrete: extracted knowledge supports neither executability checking, nor dependency verification, nor attack-goal reachability analysis.
Closing this gap requires each attack step to be modeled as an executable state transition rather than as a descriptive record, so that it can be converted into the rule form used by logic-based attack-graph engines such as MulVAL~\cite{MulVAL2005}.

\noindent\textbf{Gap 2 (Extraction).}
A straightforward approach is to prompt an LLM to generate the complete precondition--behavior--postcondition chain end to end, since LLMs have shown strong performance in CTI extraction, ATT\&CK mapping, and knowledge graph construction~\cite{Looking2023, AECR2025, CTIThinker2026, huang2024ctikg}.
However, end-to-end generation is unreliable for several reasons.
First, generated conditions are often expressed in open-ended natural language, which makes exact symbolic reasoning difficult.
Second, CTI reports rarely describe attack steps as explicit state transitions; the enabling conditions and resulting effects of a behavior are often implicit, scattered across paragraphs, or only clarified by later context.
Third, generated steps may drift in granularity, causing intermediate states to be omitted, merged, or fragmented~\cite{Survey2025}.
Finally, end-to-end generation provides no explicit mechanism to check whether each generated step is executable from the initial facts and the states produced by earlier steps.
As a result, missing conditions, unsupported intermediate states, and broken dependencies may remain hidden until formal reasoning fails.
Closing this gap requires more than prompting a stronger model. 
The extraction process must be decomposed into controllable stages that stabilize attack-step granularity,
recover enabling and resulting states, normalize them into a shared predicate space, and verify that the resulting units form a connected derivation.

This paper proposes a multi-stage framework for extracting reachable attack chains from CTI reports.
Targeting Gap~1, the framework models attack processes as state-transition chains composed of \emph{attack units}:
$\text{Attack Unit} = \langle \text{Preconditions},\; \text{Attack Behavior},\; \text{Postconditions} \rangle $.
Each attack unit represents one state-changing attack step.
The preconditions describe the environmental states, user actions, or previously derived attack states required before the behavior can occur, while the postconditions describe the new security states produced after execution.
Under this representation, an extracted attack step is not merely a textual event, a graph edge, or an ATT\&CK label, but an executable transition rule whose postconditions can satisfy the preconditions of subsequent units, enabling state matching, dependency verification, and attack-goal reachability reasoning.

Targeting Gap~2, the framework replaces single-pass generation with a controllable pipeline: attack behavior skeleton extraction, precondition and postcondition extraction, predicate normalization, diagnosis-guided repair, and Datalog-based reachability reasoning.
A behavior vocabulary constrains the granularity of extracted attack steps, a predicate vocabulary normalizes free-text conditions into unified symbolic states, and diagnosis-guided repair checks and fixes broken dependencies before the units are converted into rules.
The main contributions are as follows:
\begin{itemize}
\item \textbf{A reasoning-ready representation (Gap~1).} We model each attack step as an attack unit whose postconditions can satisfy the preconditions of subsequent units, turning a narrative report into a chain of reachable state transitions rather than descriptive IOC, entity, relation, or TTP records.

\item \textbf{A reliable extraction framework (Gap~2).} We replace end-to-end generation with a multi-stage LLM-assisted pipeline. A behavior vocabulary stabilizes attack-step granularity, a predicate vocabulary normalizes free-text conditions into a shared symbolic space, and diagnosis-guided repair enforces dependency consistency.

\item \textbf{A bridge from CTI text to formal reasoning.} Normalized attack units are compiled into Datalog-style rules, which are used for MulVAL-style reachability reasoning and path search.
These rules have traditionally required expert hand-writing; our framework produces them automatically from narrative reports.

\item \textbf{Evaluation of reasoning support.} On 20 CTI reports with 334 human-validated annotated steps, 
our framework achieves the highest annotated-step coverage among the compared systems.
It produces more complete and consistent preconditions and postconditions than end-to-end LLM generation. 
Datalog inference on the extracted chains reaches the specified attack goal in 19 of 20 reports.
\end{itemize}

The remainder of this paper is organized as follows.
Section~\ref{sec:related} reviews related work.
Section~\ref{sec:attack-unit-formulation} defines the attack-unit representation and reasoning semantics.
Section~\ref{sec:framework} presents the proposed extraction and reasoning framework.
Section~\ref{sec:evaluation} reports the experimental results.
Sections~\ref{sec:discussion} and~\ref{sec:conclusion} discuss limitations and conclude the paper.

\section{Related Work}
\label{sec:related}

\subsection{CTI Extraction and Graph Construction}

A large body of work transforms unstructured CTI reports into machine-readable knowledge, but the output remains descriptive rather than executable. 
Early systems extract IOCs---IPs, domains, hashes, and vulnerability identifiers~\cite{ChainSmith2018, Automated2020, TIMiner2020}; for example, iACE distinguishes true indicators from benign technical strings using contextual patterns~\cite{Acing2016}. 
Such indicators support detection and sharing but carry no information about dependencies among attack steps. 
A second line maps CTI text to ATT\&CK tactics, techniques, and procedures~\cite{Threat2022, TTPHunter2023, Automated2024, 259697}: TTPDrill uses a threat-action ontology~\cite{TTPDrill2017}, and later methods adopt domain-specific language models for sentence- or report-level TTP classification~\cite{TTPXHunter2024}. 
A recent SoK shows that existing TTP extraction methods are difficult to compare because they rely on different task settings, TTP ontologies, datasets, and evaluation protocols~\cite{SoK2025}.
More importantly for attack-path reasoning, these methods mainly identify which ATT\&CK techniques appear in a report, but do not model the execution conditions, resulting states, or dependencies among attack steps.
Annotated resources such as
CTI-HAL~\cite{CTIHAL2025} and AnnoCTR~\cite{lange-etal-2024-annoctr}
improve fine-grained labeling, but still target entities, concepts, or
technique labels rather than executable state dependencies.

To capture richer context, recent work builds graph-based representations from CTI reports. 
Provenance-style attack behavior graphs, such as Extractor~\cite{Extractor2021}, map system-level operations to entities such as processes, files, registries, and network sockets, making them suitable for alignment with audit logs. 
However, they primarily describe observable interactions---such as read, write, execute, and send---rather than the conditions required by each behavior and the security states it produces. 
ATT\&CK-aware knowledge graphs organize attack entities, behavioral relations, and technique-level knowledge.
AttacKG~\cite{AttacKG2022} aligns extracted attack graphs with ATT\&CK technique templates, while AttacKG+~\cite{AttacKG2025} extends this representation with behavior graphs, TTP labels, and phase-level state summaries in a multi-layer knowledge graph. 
Attack-scenario approaches focus more directly on graph completeness and plausibility.
For example, CRUcialG~\cite{CRUcialG2025} verifies the consistency of attack stages and repairs incomplete
scenario-graph structures.

LLM-assisted methods further improve graph construction and completion. 
CTINEXUS~\cite{CTINexus2025} uses optimized in-context learning, hierarchical entity alignment, and long-distance relation prediction to construct and complete entity--relation graphs. LLM-TIKG~\cite{LLMTIKG2024} and
CTI-Thinker~\cite{CTIThinker2026} similarly extract entities, relations, and TTPs for graph storage, retrieval, and question answering. 
Together, these methods represent complementary forms of CTI-derived structure, ranging from behavior graphs
and ATT\&CK-aware knowledge graphs to repaired scenario graphs and LLM-generated relation graphs.
However, these methods mainly represent relationships and ordering among entities and behaviors. Their outputs remain graph nodes, edges, relations, or stage-level summaries rather than executable rules that specify the preconditions and postconditions of each attack step.
Consequently, they cannot directly determine whether an attack behavior is executable under the current system state or whether its effects enable a subsequent step.

\subsection{Logical Attack Graph Reasoning}

Logical attack graphs provide exactly the reasoning machinery our
representation targets. MulVAL~\cite{MulVAL2005} models
vulnerabilities, host configurations, network access, and privileges in
Datalog, and derives multi-stage attack paths by inference: a rule body
encodes the conditions required for an attack action and the rule head
encodes the resulting security state, naturally capturing
precondition--postcondition dependencies. Scalable logical
attack-graph generation~\cite{Scalable2006} builds on this by encoding
causal dependencies among facts, rules, and derived privileges,
avoiding enumeration of global system states.

However, these systems assume the required facts and rules are
\emph{already} available in structured form---typically from
vulnerability scanners, configuration inventories, and expert-defined
exploit rules. They solve downstream reachability once rules exist, but
not the upstream problem of acquiring those rules from natural-language
CTI. Our work is complementary: we extract attack units from CTI
reports, normalize their preconditions and postconditions into a
predicate vocabulary, and convert them into Datalog-style rules,
thereby supplying the rules that logical attack-graph analyzers have
traditionally required experts to write by hand.

\section{Preliminaries}
\label{sec:attack-unit-formulation}

The objective of this paper is to transform unstructured CTI narratives into reachable attack chains that can be used for reachability reasoning. Instead of only identifying security entities, indicators, or ATT\&CK techniques, we focus on the state dependencies among attack behaviors: what conditions must hold before an attack step can occur, and what security states are produced after it is executed.

\subsection{Task Definition}

Given a CTI report $\mathcal{D}$, a predicate vocabulary $\mathcal{V}$, and a behavior class vocabulary $\mathcal{B}$, the extraction task aims to produce a set of normalized attack units:
\begin{equation}
\mathcal{U}=\{u_1,u_2,\ldots,u_n\}.
\end{equation}
Each attack unit represents one state-changing behavior described in the report. The predicate vocabulary $\mathcal{V}$ defines the symbolic state space used to represent preconditions and postconditions, while the behavior vocabulary $\mathcal{B}$ constrains the granularity of extracted behaviors. Given an initial fact set $F_0$ and an attack goal $g$, the reasoning task determines whether $g$ can be derived from $F_0$ using the rules converted from $\mathcal{U}$.

\subsection{Attack Unit Representation}

We model each attack step as an attack unit:
\begin{equation}
u_j=\langle Pre_j, Act_j, Post_j \rangle .
\end{equation}
Here, $Pre_j$ denotes the conditions required before the behavior can be executed, $Act_j$ denotes the attack behavior described in the report, and $Post_j$ denotes the security states produced after execution. 
To make attack units reachable, all preconditions and postconditions are normalized into predicate instances from $\mathcal{V}$:
$
\mathrm{Pre}_j \subseteq \mathcal{V}, \quad \mathrm{Post}_j \subseteq \mathcal{V}.
$
This representation allows the postconditions of one attack unit to satisfy the preconditions of another, thereby turning a narrative attack description into a chain of state transitions. The behavior vocabulary $\mathcal{B}$ does not participate in logical inference; it is used during extraction to stabilize the semantic boundary of $\mathrm{Act}_j$ and reduce over-compression or over-fragmentation of attack steps.

\subsection{Rule and Reachability Semantics}

Each normalized attack unit
$u_j=\langle \mathit{Pre}_j,\mathit{Act}_j,\mathit{Post}_j\rangle$
defines an executable state transition: when all states in
$\mathit{Pre}_j$ hold, the behavior $\mathit{Act}_j$ can be executed and
produces the states in $\mathit{Post}_j$.

For Datalog-style reasoning, this transition is compiled into rules.
Given $\mathit{Pre}_j=\{p_1,p_2,\dots,p_m\}$ and
$\mathit{Post}_j=\{q_1,q_2,\dots,q_s\}$, where $m$ and $s$ denote
the numbers of preconditions and postconditions, respectively, we generate
one rule for each postcondition:
\begin{equation}
\label{eq:rulegen}
r_{j,l}: \quad
q_l \leftarrow p_1,p_2,\dots,p_m,
\qquad l=1,\dots,s .
\end{equation}
The rule body represents the states required for the corresponding
attack behavior to be executable, and the rule head represents one
security state produced after execution.

Let $\mathcal{R}$ be the set of generated rules. 
Starting from the initial fact set $F_0$, reachability reasoning repeatedly applies rules whose bodies are satisfied and adds their heads as new facts until the goal $g$ is reached.

This rule formulation also provides a diagnosis signal. A precondition
$p\in\mathit{Pre}_j$ is \emph{unsupported} if it is neither an initial
fact nor produced by any preceding attack unit:
\begin{equation}
\label{eq:unsupported}
p \notin F_0
\quad\text{and}\quad
p \notin \bigcup_{k<j} \mathrm{Post}_k .
\end{equation}
Unsupported preconditions indicate potential defects such as missing
initial facts, missing intermediate states, misplaced pre/postconditions,
or inconsistent predicate normalization.

\section{Framework}
\label{sec:framework}

This section presents the proposed framework for extracting reachable attack chains from CTI reports. The framework transforms unstructured threat narratives into normalized attack units and further converts them into Datalog-style rules for reachability reasoning.

\subsection{Overview}
\label{subsec:framework_overview}

\begin{figure*}[!t]
    \centering

    \begin{subfigure}[t]{\textwidth}
        \centering
        \includegraphics[width=\linewidth]{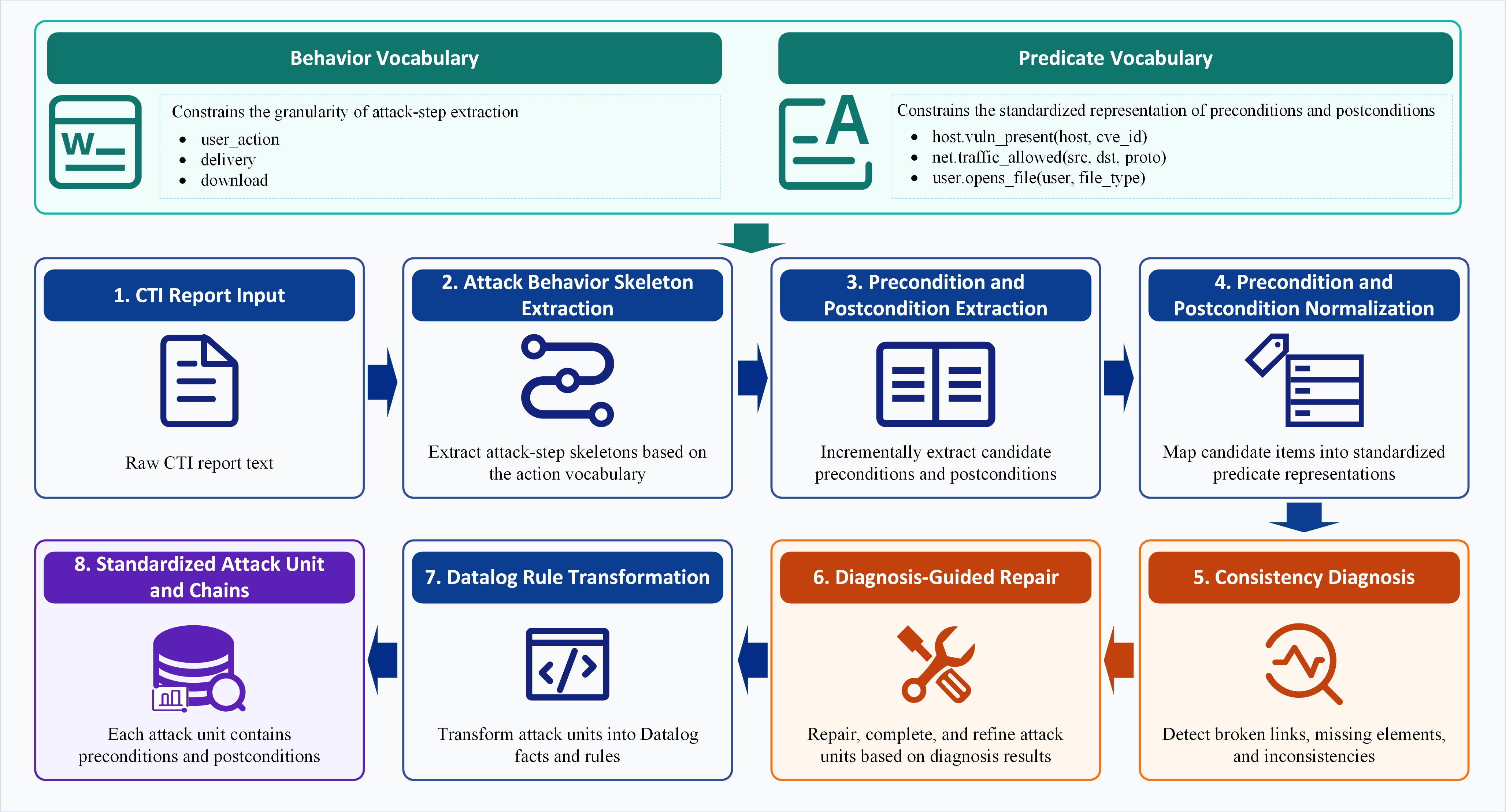}
        \caption{CTI-to-Attack-Unit Repository Construction.}
        \label{fig:overview_stage1}
    \end{subfigure}

    \vspace{0.8em}

    \begin{subfigure}[t]{\textwidth}
        \centering
        \includegraphics[width=\linewidth]{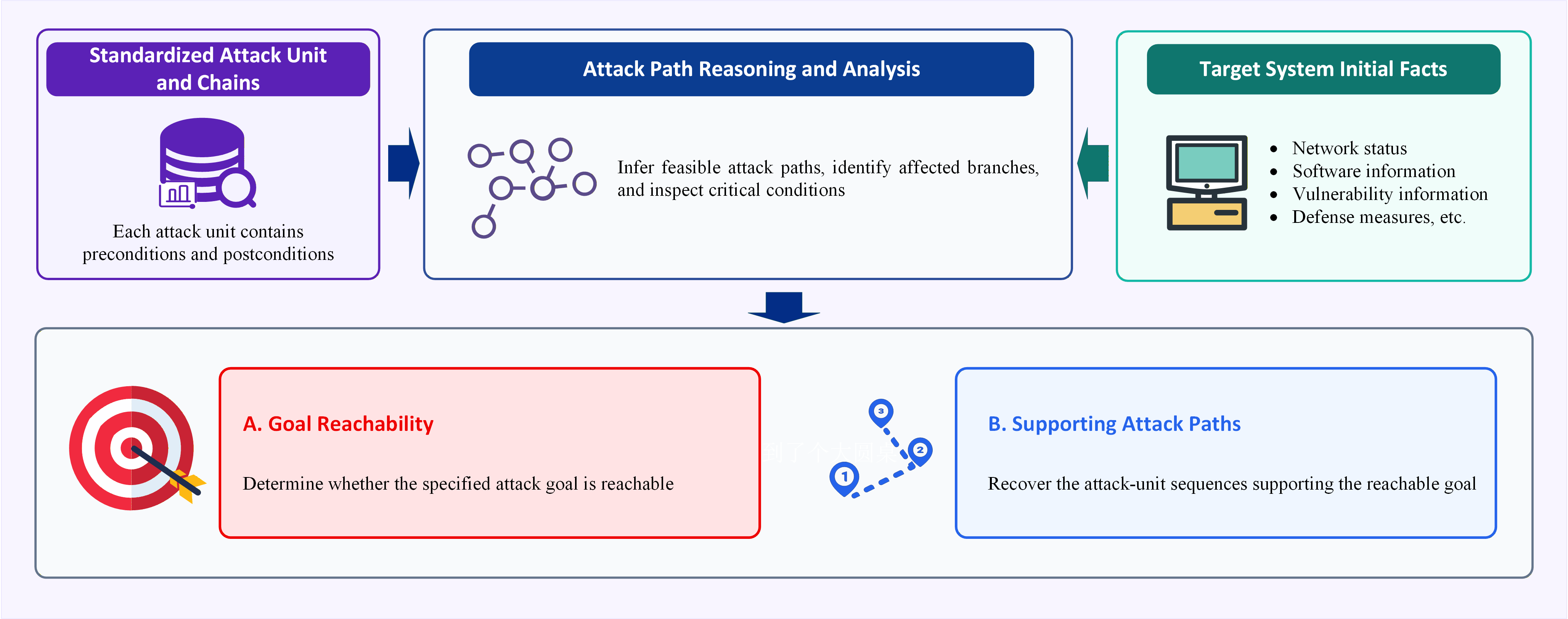}
        \caption{Repository-Based Attack Path Reasoning.}
        \label{fig:overview_stage2}
    \end{subfigure}

    \caption{Overview of the proposed CTI-to-Attack-Unit extraction and attack path reasoning framework.}
    \label{fig:overview}
\end{figure*}

As shown in Figure~\ref{fig:overview}, the framework operates in two phases.
The first phase, CTI-to-Attack-Unit Repository Construction, transforms raw CTI reports into reasoning-ready attack knowledge. 
It extracts attack behavior skeletons, recovers and normalizes preconditions and postconditions, repairs inconsistent dependencies, and converts the standardized attack units into Datalog-style rules.

The second phase, Repository-Based Attack Path Reasoning, uses the generated attack-unit repository together with a target system's initial facts to perform reachability reasoning. It determines whether the specified attack goal is reachable and recovers the supporting attack paths.

The framework relies on two vocabularies. The \emph{behavior vocabulary} (informed by MITRE ATT\&CK
tactics~\cite{mitre2026attackenterprisea,strom2020mitreattacka} and common CTI action semantics) constrains the granularity of attack-step extraction, and the \emph{predicate vocabulary} (following MulVAL-style~\cite{MulVAL2005} precondition--postcondition semantics) normalizes free-text conditions into symbolic states. 
Both vocabularies were constructed before evaluation. 
We first built an initial version based on standard CTI knowledge, MITRE ATT\&CK~\cite{strom2020mitreattacka,mitre2026attackenterprisea}. 
We then refined the vocabularies with five CTI reports outside the evaluation dataset.
Appendix~\ref{app:behavior_vocab} and Appendix~\ref{app:predict_vocab} provides the templates and construction details.

\subsection{Attack Behavior Skeleton Extraction}
\label{subsec:skeleton_extraction}
CTI reports usually contain mixed information, including attack procedures, vulnerability analysis, malware implementation details, indicators of compromise, and mitigation suggestions. Not all such information corresponds to attack-chain steps. The goal of attack behavior skeleton extraction is therefore to identify an ordered sequence of attack behaviors that directly participate in the observed attack process.

Given a CTI report $D_i$, the framework extracts an ordered attack behavior skeleton:
\begin{equation}
A_i = \langle \mathrm{Act}^i_1,\ldots,\mathrm{Act}^i_{n_i}\rangle.
\end{equation}
Each $\mathrm{Act}^i_j$ describes one core attack behavior, such as opening a malicious document, downloading a payload, executing a script, bypassing a defense mechanism, or establishing a C2 channel.

A major challenge in this stage is inconsistent step granularity: overly coarse steps merge multiple state-changing behaviors, making it difficult to assign accurate pre- and postconditions, while overly fine steps fragment the chain into actions that are not meaningful state transitions. 
We therefore use the behavior vocabulary $\mathcal{B}$ as a granularity anchor. 
It contains 16 coarse-grained classes such as \texttt{user\_action}, \texttt{download}, \texttt{execute}, and \texttt{communication}. 
During skeleton extraction, each behavior is assigned one class to keep the 
step boundaries consistent. 
For example, the phrase ``the victim opens a malicious PDF and downloads an RTF document'' is split into a \texttt{user\_action} step and a \texttt{download} step.

\subsection{Precondition and Postcondition Extraction}
\label{subsec:pre_post_extraction}

After obtaining the attack behavior skeleton, the framework extracts
the preconditions and postconditions of each attack behavior. For a
behavior $Act^i_j$, the LLM generates candidate preconditions
$Pre^{i,\mathrm{cand}}_j$ and postconditions $Post^{i,\mathrm{cand}}_j$:
preconditions describe the states that must hold before the behavior
can occur, and postconditions describe the states produced after the behavior is executed.

To improve coverage while limiting context noise, we use a
\emph{dual-channel} strategy. The \emph{local channel} feeds the LLM
the current step, its neighboring steps, and a report fragment
retrieved around relevant keywords, focusing on local textual evidence
and suppressing irrelevant context. The \emph{global channel} feeds the
complete CTI report, allowing the model to capture long-range
dependencies that span multiple paragraphs.
For each behavior, the candidates from the two channels are then fused
into a single set of preconditions and postconditions. The fusion
prioritizes conditions supported by both channels, merges
near-duplicates, drops unsupported or overly general items, and
reassigns conditions that are attributed to the wrong neighboring step.
To avoid noisy expansion, each step retains only a small number of
high-confidence preconditions and postconditions.

\subsection{Predicate Normalization}
\label{subsec:predicate_normalization}

The fused preconditions and postconditions are still natural-language phrases, whereas reachability reasoning requires symbolic states that can be matched exactly across attack units. 
The normalization stage therefore maps each candidate condition $c$ to one or more predicate instances from the predefined predicate vocabulary $\mathcal{V}$:
\begin{equation}
Normalize(c, \mathcal{V}) = \{p_1, p_2, \ldots, p_k\},
\quad p_l \in \mathcal{V}.
\end{equation}
where each $p_l$ is a predicate instance representing one security
state described by $c$.
Each predicate instance takes the form \texttt{category.predicate\_name(param1, param2, ...)}: the category classifies the type of security state, the predicate name specifies the state property or relation, and the parameters bind the entities involved. 
For example, ``the victim host is vulnerable to CVE-2017-11882'' is normalized as \texttt{host.vuln\_present(victim\_host,
CVE-2017-11882)}, and ``the malware establishes an HTTPS C2 channel''
as \texttt{data.c2\_channel( victim\_host, c2\_server, HTTPS)}.

We apply three simple normalization rules. 
First, each condition must be mapped to predicates in the predicate vocabulary $\mathcal{V}$, rather than to newly invented predicate names. 
Second, a compound condition is split into multiple predicates when it contains multiple states. 
Third, the same entity is written with the same canonical name across the chain. 
These rules help the postconditions of one unit match the preconditions of later units.

After normalization, each attack unit becomes a fully symbolic state
transition
\begin{equation}
u^i_j = \langle NormPre^i_j,\, \mathrm{Act}^i_j,\, NormPost^i_j \rangle,
\end{equation}
with $\mathrm{NormPre}^i_j, \mathrm{NormPost}^i_j \subseteq \mathcal{V}$, ready for
consistency diagnosis and rule generation. 
Representative predicate templates are listed in Appendix~\ref{app:predict_vocab}.

\subsection{Diagnosis-Guided Repair}
\label{subsec:diagnosis_repair}

After normalization, some attack units may still contain errors that prevent the chain from being used for reachability reasoning. 
For example, a precondition may not be produced by any earlier step, or a predicate may
not match the meaning of the report. 
We therefore add a diagnosis and repair stage before rule generation.

The diagnosis stage uses an LLM to inspect the normalized attack chain and produce structured descriptions of potential inconsistencies and corresponding repair suggestions. The diagnosed problems may involve unsupported dependencies, predicates inconsistent with the report evidence, or incorrectly represented branch relationships.
First, it checks whether the chain is connected. Each
precondition should either be produced by an earlier attack
unit or be included in the initial fact set $F_0$. 
Second, it checks whether each normalized predicate still matches the report evidence. 
Third, it checks whether alternative attack branches have been incorrectly merged into one path. 
Each detected problem is recorded with an issue type and severity.
The repair stage strictly follows the diagnosis results. It only fixes the
flagged issues and does not add, delete, or rewrite other parts of the
chain. We run this process twice: the first round applies the fixes, and the
second round checks whether any flagged issues remain. Appendix~\ref{app:diagnosis}
lists all issue types and repair actions.

\subsection{Datalog Rule Generation}
\label{subsec:datalog_generation}

After diagnosis and repair, we convert each normalized attack unit into Datalog-style rules. 
The normalized preconditions become the rule body, and each normalized postcondition becomes a rule head. 
When an attack unit has multiple postconditions, we generate one rule for each postcondition, following Eq.~(\ref{eq:rulegen}). 
Each rule is stored as $r = \langle rule\_id, head, body \rangle$. 
The complete set of rules forms the attack rule set $\mathcal{R}$ used for downstream reachability reasoning.

\subsection{Attack Path Reasoning}
\label{subsec:attack_path_reasoning}

Given the rule set $\mathcal{R}$, the initial fact set $F_0$, and the attack goal $g$, the framework first checks whether $g$ is reachable.
Starting from the initial fact set $F_0$, reachability reasoning repeatedly applies satisfied rules and adds their heads as new facts, until either the goal $g$ is derived or no new facts can be derived.

For reachable goals, the framework traces the derivation backward from $g$ to the initial facts, producing a set of attack paths $Paths(g)$. 
Each path is an ordered list of attack units that contribute to the goal.

\section{Evaluation}
\label{sec:evaluation}

\subsection{Experimental Setup}
\label{sec:setup}

\subsubsection{Dataset and Ground Truth} 
We evaluate the proposed framework on 20 publicly available CTI reports~\cite{Mining2025, AttacKG2022}, containing 334 executable annotated attack steps. 
The reports cover diverse attack scenarios, ranging from spear phishing, watering-hole attacks, and document exploitation to anti-analysis checks, DLL search-order hijacking,
memory exploitation, and C2 communication.
System-initial-condition records, such as software versions, vulnerability existence, binary availability, and network reachability, are used only as candidate initial facts for Datalog reasoning and are not counted as executable attack steps.

The ground truth is constructed with an LLM-assisted annotation and manual review process. 
We first use \texttt{gpt-5.5}~\cite{gpt55} to identify candidate state-changing attack behaviors from each report, and then verify the candidates with \texttt{Claude Opus 4.8}~\cite{claudeopus48} and manual review.
Each annotated step corresponds to one core state-changing attack behavior and is represented in structured JSON with \texttt{pre\_groups} and \texttt{post\_groups}.
Predicates within the same group represent equivalent or alternative expressions, while different groups represent separate required conditions or effects.

All 20 reports are used for attack-step coverage and predicate-level evaluation. Among them, four representative reports are further used for detailed case analysis and ablation: DeputyDog, Frankenstein, OceanLotus, and Cobalt~\cite{AttacKG2022}. 
These reports cover different combinations of report length, chain topology, and branching structure.

\subsubsection{Compared Methods and Implementation}

We compare our framework with three representative CTI
extraction systems: \textbf{AttacKG+}~\cite{AttacKG2025}, \textbf{CRUcialG}~\cite{CRUcialG2025}, and \textbf{CTINEXUS}~\cite{CTINexus2025}. 
These methods produce heterogeneous outputs, such as entity-relation triples, attack graphs, TTP labels, or cyber threat knowledge graphs. 
Since they are not designed to output explicit precondition--postcondition attack units, directly comparing predicate-level recall would be unfair. 
We therefore compare all methods using attack-step coverage, which evaluates whether the core attack behaviors in the annotated steps are recovered.

Our framework uses \texttt{gpt-4.1}~\cite{gpt41} for attack behavior skeleton extraction, pre/postcondition extraction, and predicate normalization, and uses \texttt{gpt-5.2}~\cite{gpt52} for diagnosis-guided repair.
Full prompts and decoding parameters are provided in our code\footnote{\url{https://anonymous.4open.science/r/artifact064b}}. 
The Datalog reasoning stage uses no LLM. 
The internal ablation variants are described in Section~\ref{sec:ablation}.

\paragraph{Reasoning configuration}
In practical deployment, $F_0$ would be provided by the configuration of the target system. Since the evaluated CTI reports are not accompanied by concrete target-system inventories, we construct $F_0$ from static conditions supported or reasonably implied by each report,
such as vulnerability presence, software or binary availability, credentials, and network connectivity. 
States produced by attacker behaviors, including downloaded payloads, started processes, and
established C2 channels, are excluded from $F_0$ and must be derived from the extracted rules.

For each report, we select a security-relevant terminal state described in the report as the reasoning goal. 
The purpose of this setup is to evaluate the internal reasoning readiness of the extracted chain: whether its attack units form a connected derivation from the report-supported initial conditions to the selected terminal state.

\subsubsection{Evaluation Metrics}
\label{subsubsec:metrics}

We use two complementary evaluation metrics. 

First, we use \emph{attack-step coverage} to compare CTI extraction systems.
Because the methods produce different types of results, we manually align their extractions with the annotated attack steps.
The alignment is ground-truth-step-centered: one ground-truth step may align with one extracted unit or multiple finer-grained extracted units, but it contributes only one score. 
Each ground-truth step is labeled \emph{Full}, \emph{Partial}, or \emph{Miss} according to its coverage. 
A \emph{Full} match requires the aligned extraction to recover the core attack behavior and the essential behavioral details needed to distinguish the step, such as its target, object, or execution mechanism. 
A \emph{Partial} match is assigned when the core behavior is recognizable but one or more distinguishing
details are omitted, over-generalized, or only indirectly represented. 
A \emph{Miss} indicates that no extracted output can be semantically aligned with the annotated step. 
For each ground-truth step $s$, we define its coverage score as:
\begin{equation}
\operatorname{score}(s)=
\begin{cases}
1,   & \text{Full},\\
0.5, & \text{Partial},\\
0,   & \text{Miss}.
\end{cases}
\end{equation}

We report the numbers of steps labeled \emph{Full}, \emph{Partial}, and \emph{Miss} (F/P/M), together with
\emph{Step Coverage}. Step Coverage is the mean coverage score over all ground-truth executable steps:
\begin{equation}
\label{eq:step_coverage}
\mathrm{StepCoverage}
=
\frac{1}{N}\sum_{s=1}^{N}\operatorname{score}(s)
=
\frac{F+0.5P}{F+P+M}
\times 100\%,
\end{equation}
where $N=F+P+M$ is the number of ground-truth executable
steps.

Second, we use \emph{pre/postcondition coverage} to evaluate whether the extracted preconditions and postconditions match the ground-truth annotation.
This metric is used only for our framework and its internal variants, because the compared systems do not produce normalized precondition--postcondition attack units.
An extracted predicate is counted as correct only when its predicate and normalized parameters match those in the ground-truth annotation.
In some cases, the same condition or effect may have multiple valid predicate expressions. 
We put these expressions into the same alternative group. 
Matching any predicate in the group counts as covering that condition or effect, and the group is counted only once.

We report Predicate F1 across all precondition and postcondition items, along with Precondition Recall and Postcondition Recall.
Recall measures the fraction of ground-truth items covered by the extraction, and precision measures the fraction of extracted items supported by the ground truth.

\subsection{Attack-Step Coverage}
\label{subsec:rq1}

\begin{figure*}[t]
\centering
\begin{subfigure}[t]{0.49\textwidth}
     \centering
     \includegraphics[width=\linewidth]{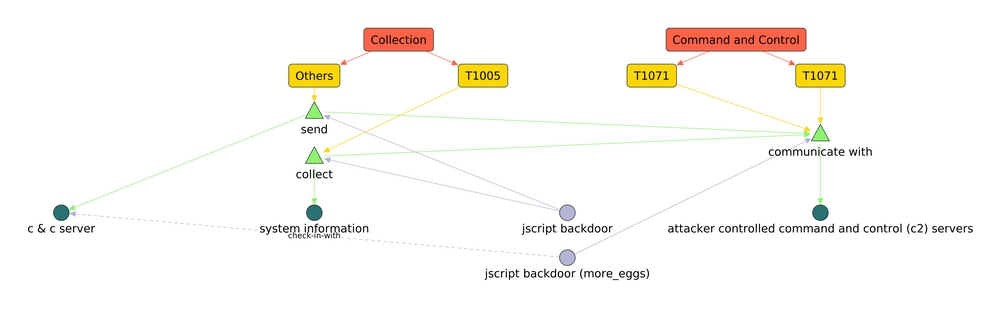}
     \caption{AttacKG+}
\end{subfigure}
\hfill
\begin{subfigure}[t]{0.49\textwidth}
     \centering
     \includegraphics[width=\linewidth]{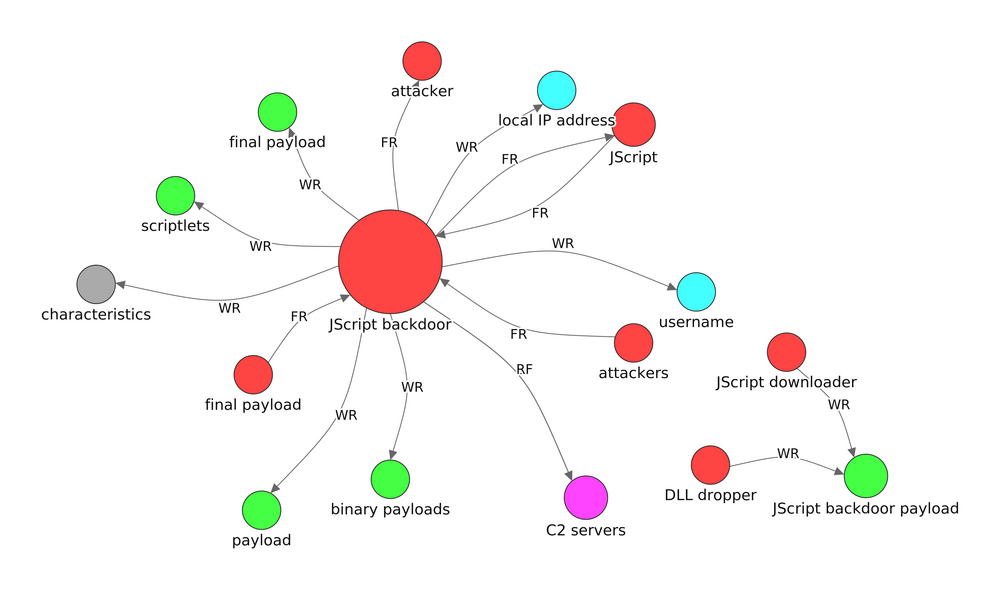}
     \caption{CRUcialG}
\end{subfigure}
\begin{subfigure}[t]{0.49\textwidth}
    \centering
    \includegraphics[width=\linewidth]{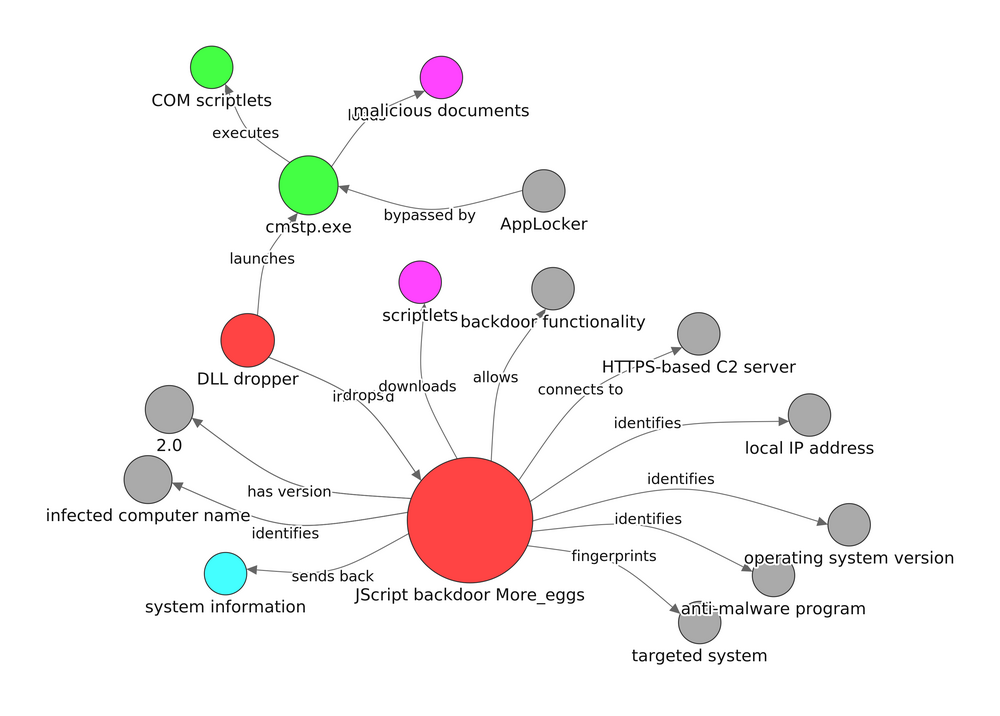}
    \caption{CTINEXUS}
\end{subfigure}
\hfill
\begin{subfigure}[t]{0.49\textwidth}
    \centering
    \includegraphics[width=\linewidth]{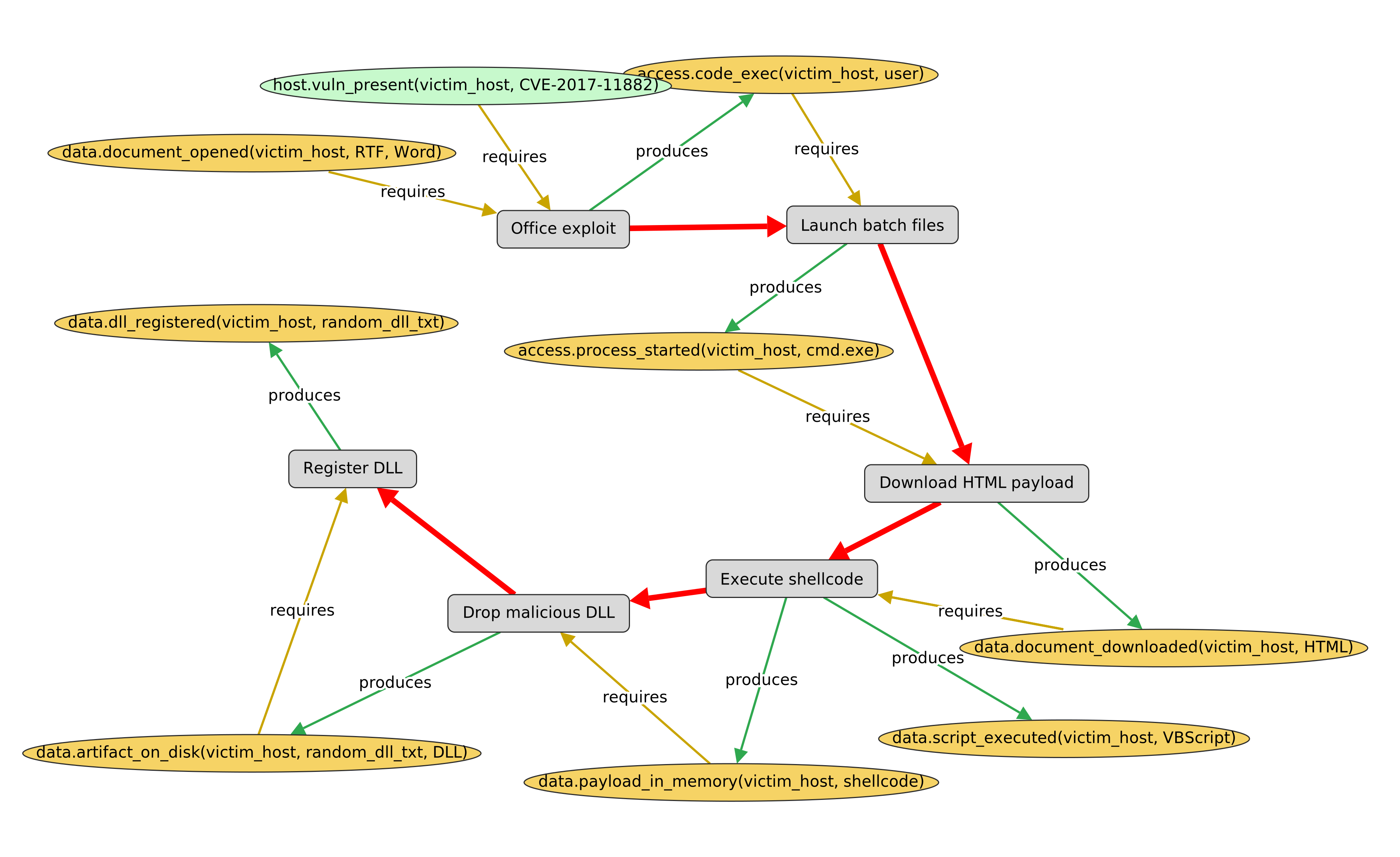}
    \caption{Ours}
\end{subfigure}

\caption{Local comparison of different extraction methods on the Cobalt report. The selected region corresponds to the JScript backdoor branch, where the proposed framework preserves explicit state transitions between adjacent attack steps.}
\label{fig:local_comparison}
\end{figure*}

Figure~\ref{fig:local_comparison} shows a local example from the Cobalt JScript-backdoor branch. 
The compared methods recover different kinds of information: AttacKG+ extracts entities
and high-level relations, CRUcialG builds a coarse scenario graph, and CTINEXUS produces relation triples. 
In contrast, our framework extracts attack units with explicit preconditions and postconditions, making the state transitions between adjacent steps visible.

\begin{table*}[t]
\centering
\caption{Attack-step coverage comparison with existing CTI
extraction methods on 20 CTI reports. For each method, F/P/M
denotes the numbers of annotated steps that are fully covered,
partially covered, and missed, respectively. The best Step
Coverage result for each report is highlighted in bold.}
\label{tab:external_step_coverage}
\setlength{\tabcolsep}{3.0pt}
\begin{tabular}{@{}l c cc cc cc cc@{}}
\toprule
Report
& Ground Truth
& \multicolumn{2}{c}{AttacKG+}
& \multicolumn{2}{c}{CRUcialG}
& \multicolumn{2}{c}{CTINEXUS}
& \multicolumn{2}{c}{\textbf{Ours}} \\
\cmidrule(lr){3-4}
\cmidrule(lr){5-6}
\cmidrule(lr){7-8}
\cmidrule(lr){9-10}
&
& \StepFPMHead &  Coverage(\%)
& \StepFPMHead &  Coverage(\%)
& \StepFPMHead &  Coverage(\%)
& \StepFPMHead &  Coverage(\%) \\
\midrule
Cobalt
& 27
& \StepFPM{6}{20}{1}    & 59.3
& \StepFPM{6}{19}{2}    & 57.4
& \StepFPM{18}{8}{1}    & 81.5
& \StepFPM{23}{4}{0}    & \textbf{92.6} \\

DeputyDog
& 12
& \StepFPM{5}{7}{0}     & 70.8
& \StepFPM{3}{6}{3}     & 50.0
& \StepFPM{9}{3}{0}     & 87.5
& \StepFPM{11}{1}{0}    & \textbf{95.8} \\

Frankenstein
& 19
& \StepFPM{14}{5}{0}    & 86.8
& \StepFPM{3}{11}{5}    & 44.7
& \StepFPM{16}{3}{0}    & \textbf{92.1}
& \StepFPM{16}{3}{0}    & \textbf{92.1} \\

OceanLotus
& 15
& \StepFPM{9}{6}{0}     & 80.0
& \StepFPM{3}{7}{5}     & 43.3
& \StepFPM{9}{6}{0}     & 80.0
& \StepFPM{12}{3}{0}    & \textbf{90.0} \\

XAgentOSX
& 17
& \StepFPM{13}{4}{0}    & \textbf{88.2}
& \StepFPM{2}{10}{5}    & 41.2
& \StepFPM{13}{4}{0}    & \textbf{88.2}
& \StepFPM{13}{4}{0}    & \textbf{88.2} \\

macOS.Macma
& 13
& \StepFPM{8}{4}{1}     & 76.9
& \StepFPM{4}{3}{6}     & 42.3
& \StepFPM{9}{3}{1}     & 80.8
& \StepFPM{10}{2}{1}    & \textbf{84.6} \\

Emotet Campaigns
& 12
& \StepFPM{8}{4}{0}     & 83.3
& \StepFPM{5}{5}{2}     & 62.5
& \StepFPM{10}{2}{0}    & 91.7
& \StepFPM{12}{0}{0}    & \textbf{100.0} \\

Silence Trojan
& 18
& \StepFPM{11}{7}{0}    & 80.6
& \StepFPM{7}{11}{0}    & 69.4
& \StepFPM{12}{6}{0}    & 83.3
& \StepFPM{18}{0}{0}    & \textbf{100.0} \\

Elfin / APT33
& 23
& \StepFPM{18}{3}{2}    & 84.8
& \StepFPM{10}{10}{3}   & 65.2
& \StepFPM{15}{6}{2}    & 78.3
& \StepFPM{22}{1}{0}    & \textbf{97.8} \\

SpeakUp
& 21
& \StepFPM{16}{5}{0}    & 88.1
& \StepFPM{2}{6}{13}    & 23.8
& \StepFPM{10}{5}{6}    & 59.5
& \StepFPM{18}{3}{0}    & \textbf{92.9} \\

TSCookie
& 15
& \StepFPM{11}{4}{0}    & 86.7
& \StepFPM{1}{6}{8}     & 26.7
& \StepFPM{11}{4}{0}    & 86.7
& \StepFPM{15}{0}{0}    & \textbf{100.0} \\

Sodinokibi Crescendo
& 27
& \StepFPM{18}{8}{1}    & 81.5
& \StepFPM{1}{9}{17}    & 20.4
& \StepFPM{13}{8}{6}    & 63.0
& \StepFPM{22}{5}{0}    & \textbf{90.7} \\

Sofacy Fysbis
& 12
& \StepFPM{6}{6}{0}     & 75.0
& \StepFPM{0}{6}{6}     & 25.0
& \StepFPM{5}{5}{2}     & 62.5
& \StepFPM{10}{1}{1}    & \textbf{87.5} \\

Gamaredon COVID-19
& 12
& \StepFPM{8}{4}{0}     & 83.3
& \StepFPM{0}{6}{6}     & 25.0
& \StepFPM{11}{1}{0}    & 95.8
& \StepFPM{12}{0}{0}    & \textbf{100.0} \\

BabyShark
& 17
& \StepFPM{10}{7}{0}    & 79.4
& \StepFPM{1}{14}{2}    & 47.1
& \StepFPM{11}{6}{0}    & 82.4
& \StepFPM{17}{0}{0}    & \textbf{100.0} \\

Deep Panda
& 15
& \StepFPM{11}{4}{0}    & 86.7
& \StepFPM{0}{11}{4}    & 36.7
& \StepFPM{13}{2}{0}    & 93.3
& \StepFPM{15}{0}{0}    & \textbf{100.0} \\

HermeticWiper
& 17
& \StepFPM{10}{7}{0}    & 79.4
& \StepFPM{1}{5}{11}    & 20.6
& \StepFPM{9}{8}{0}     & 76.5
& \StepFPM{12}{5}{0}    & \textbf{85.3} \\

Rover Trojan
& 16
& \StepFPM{13}{3}{0}    & 90.6
& \StepFPM{6}{8}{2}     & 62.5
& \StepFPM{14}{2}{0}    & 93.8
& \StepFPM{16}{0}{0}    & \textbf{100.0} \\

DEV-0586 / WhisperGate
& 8
& \StepFPM{5}{3}{0}     & 81.2
& \StepFPM{1}{6}{1}     & 50.0
& \StepFPM{6}{2}{0}     & 87.5
& \StepFPM{8}{0}{0}     & \textbf{100.0} \\

Sednit Zebrocy
& 18
& \StepFPM{12}{5}{1}    & 80.6
& \StepFPM{4}{6}{8}     & 38.9
& \StepFPM{12}{5}{1}    & 80.6
& \StepFPM{18}{0}{0}    & \textbf{100.0} \\
\midrule
Total F/P/M
& 334
& \StepFPM{212}{116}{6} & -
& \StepFPM{60}{165}{109} & -
& \StepFPM{226}{89}{19} & -
& \StepFPM{300}{32}{2} & - \\
Average
& -
& - & 81.1
& - & 42.6
& - & 82.2
& -  & \textbf{94.9} \\
\bottomrule
\end{tabular}
\end{table*}

Because the compared methods do not generate normalized precondition--postcondition units, we evaluate them at the attack-step level. 
Attack-step coverage measures how well the extracted steps cover the ground-truth attack steps.
Table~\ref{tab:external_step_coverage} reports attack-step coverage on 20 reports containing 334 annotated executable steps. 
Our framework achieves the highest macro-average Step Coverage of 94.9\%, compared with 82.2\% for CTINEXUS,
81.1\% for AttacKG+, and 42.6\% for CRUcialG. 
Across all 334 annotated steps, our framework fully covers 300, partially covers 32, and misses only two. AttacKG+ and CTINEXUS achieve lower coverage primarily because more annotated steps are only partially covered, while CRUcialG misses 109 steps. 
These results show that the improvement mainly comes from recovering complete executable attack steps rather than outputs that capture only some of the relevant entities, relations, techniques, or scenario nodes.
Our framework achieves the best or tied-best result on every report, providing a strong basis for the subsequent evaluation of preconditions, postconditions, and reasoning readiness.

\subsection{Pre/Postcondition Coverage}
\label{subsec:rq2}
We evaluate pre/postcondition coverage using the predicate-level metrics defined in Section~\ref{subsubsec:metrics}.

\begin{table*}[t]
\centering
\caption{Pre/postcondition coverage on 20 CTI reports under exact
predicate checking after normalization. Predicate F1, Precision (P),
and Recall (R) are computed over all precondition and postcondition
items. Pre R and Post R report recall for preconditions and
postconditions, respectively.}
\label{tab:predicate}
\setlength{\tabcolsep}{10pt}
\begin{tabular}{lccccc}
\toprule
Report & F1(\%) & P(\%) & R(\%) & Pre R(\%) & Post R(\%) \\
\midrule
Cobalt       	& 95.7 & 95.0 & 96.3 & 94.7 & 97.7 \\
DeputyDog    	& 93.8 & 100.0 & 88.3 & 88.9 & 87.9 \\
Frankenstein 	& 93.3 & 100.0 & 87.5 & 89.3 & 86.1 \\
OceanLotus 		& 95.3 & 100.0 & 91.1 & 90.0 & 92.0 \\
XAgentOSX 		& 52.7 & 66.0 & 43.9 & 33.3 & 58.3 \\
macOS.Macma 	& 51.3 & 49.1 & 53.8 & 55.0 & 52.6 \\
Emotet Campaigns & 81.1 & 71.2 & 94.3 & 94.4 & 94.1 \\
Silence Trojan 	& 75.4 & 89.1 & 65.4 & 75.0 & 57.1 \\
Elfin / APT33 	& 81.7 & 87.9 & 76.4 & 82.1 & 69.7 \\
SpeakUp 		& 93.0 & 98.8 & 88.0 & 93.0 & 82.5 \\
TSCookie 		& 84.6 & 84.1 & 85.1 & 90.0 & 81.5 \\
Sodinokibi Crescendo & 79.1 & 76.7 & 81.6 & 90.9 & 72.1 \\
Sofacy Fysbis 	& 85.5 & 97.4 & 76.2 & 91.3 & 57.9 \\
Gamaredon COVID-19 & 71.9 & 75.7 & 68.4 & 70.0 & 66.7 \\
BabyShark 		& 77.3 & 84.6 & 71.2 & 86.2 & 56.7 \\
Deep Panda 		& 94.2 & 94.3 & 94.1 & 100.0 & 88.5 \\
HermeticWiper 	& 66.7 & 80.6 & 56.9 & 57.7 & 56.0 \\
Rover Trojan 	& 66.7 & 61.3 & 73.2 & 69.0 & 77.8 \\
DEV-0586 / WhisperGate & 91.9 & 100.0 & 85.0 & 90.9 & 77.8 \\
Sednit Zebrocy 	& 85.8 & 93.3 & 79.4 & 80.0 & 78.9 \\
\midrule
Average 		& 80.9 & 85.3 & 77.8 & 81.1 & 74.6 \\
\bottomrule
\end{tabular}
\end{table*}

Table~\ref{tab:predicate} reports the predicate-level results on the 20 CTI reports. 
The performance is consistently high for most reports. 
Twelve of the 20 reports achieve Predicate F1 above 80\%, including seven above
90\%. 
In particular, Cobalt, DeputyDog, Frankenstein, OceanLotus, SpeakUp, Deep Panda, and DEV-0586/WhisperGate all achieve Predicate F1 above 90\%. 
Averaged across the 20 reports, the framework achieves 80.9\% Predicate F1, with 85.3\% precision and 77.8\% recall overall, and precondition and postcondition recall of 81.1\% and 74.6\%, respectively. 
These results indicate that the extracted attack units generally recover both the conditions required to execute an attack behavior and the security states produced after its execution. 

To understand the failure modes of exact predicate matching, we
analyze the two lowest-scoring reports. In both, our method recovers
the underlying attack behaviors, but the resulting predicates do not
align with the annotation under exact checking.

For XAgentOSX, the low coverage comes from a coarser normalization that stays
internally consistent but no longer matches the annotated predicates. 
This mismatch comes mainly from two recurring substitutions. 
First, the running malware is represented by the generic state
\texttt{access.session(victim\_host, C2\_agent)} and reused as a precondition across almost every downstream step. 
The ground truth instead keeps two distinct states: \texttt{access.process\_started} for the running process and \texttt{access.remote\_control} for attacker control. 
Second, distinct C2 commands are each specified in the ground truth with a specific parameter, such as \texttt{c2\_command\_received(victim\_host, remote\_shell\_command)}. 
The pipeline assigns all of them the same generic parameter \texttt{attacker\_command}.

For macOS.Macma, the low coverage comes from errors of a different kind: a structural mistake in the branch topology and pervasive naming inconsistency. 
The report describes two independent delivery branches that later converge. 
One is a DMG branch that drops \texttt{UserAgent\_2019} without any exploit; the other is a vulnerability branch in which a zero-day and an N-day drop a standalone installer that installs \texttt{UserAgent\_2021}.
The pipeline collapses these into one: it models the two vulnerabilities as variants but has both produce
UserAgent\_2019, and does not recover the separate 2021 installer or the UserAgent\_2021 binary. 
As a result, the 2021 branch is not preserved as a distinct state-transition sequence, causing its associated predicate groups to remain unmatched or only partially matched under exact checking.
The remaining loss is naming drift---the same entity written as \texttt{UserAgent\_2019} versus \texttt{UserAgent 2019}, or as a canonical symbol versus a full path---which stays consistent within the chain but fails exact matching.

The unmatched predicates reveal that extraction can still suffer from over-general normalization, branch conflation, and entity-naming drift. 
Nonetheless, the framework produces accurate precondition--postcondition units on most of the CTI reports.

\subsection{Datalog Reasoning}
\label{subsec:rq3}

We evaluate whether the final attack chains can support Datalog-based reachability reasoning and attack-path recovery.
As described in Section~\ref{subsec:datalog_generation}, the normalized attack units are converted into Datalog-style rules.
For each report, we specify an attack goal. 
Forward chaining starts from the initial fact set $F_0$ and checks whether the goal can be derived. 
For a reachable goal, backward search traces the dependencies from the goal to the initial facts and recovers the supporting attack paths.

The reasoning evaluation operates on the final attack units rather than directly on the annotated attack steps. 
The number of final units may therefore exceed the number of annotated steps because one
annotated behavior may be decomposed into several finer-grained state transitions. 
For example, the 27 annotated steps in Cobalt are represented as 41 attack units. 

Table~\ref{tab:rq3} reports the reasoning results on all 20 reports.
In addition to goal reachability and the number of recovered paths, we use two deterministic consistency checks on the final chains.
Resid.\ counts preconditions that are neither included in $F_0$ nor produced by an earlier attack unit. 
NoPost counts units whose postcondition set is empty.

\begin{table*}[t]
\centering
\caption{Datalog reasoning results on the 20 CTI reports.
\#Units is the number of attack units in the final chain.
Reach indicates whether forward chaining derives the specified attack
goal from $F_0$. \#Paths is the number of supporting attack paths
recovered by backward search. Resid.\ counts preconditions that remain
unsupported in the final chain. NoPost counts units with an empty
postcondition set.}
\label{tab:rq3}
\setlength{\tabcolsep}{10pt}
\renewcommand{\arraystretch}{1.00}
\begin{tabular}{l r c r r r}
\toprule
Report & \#Units & Reach & \#Paths & Resid. & NoPost \\
\midrule
Cobalt                            & 41 & \cmark & 2 & 0 & 0 \\
DeputyDog                         & 17 & \cmark & 1 & 0 & 1 \\
Frankenstein                      & 20 & \cmark & 1 & 0 & 0 \\
OceanLotus                        & 23 & \cmark & 1 & 0 & 4 \\
XAgentOSX                         & 15 & \cmark & 1 & 0 & 0 \\
macOS.Macma                       & 17 & \cmark & 7 & 0 & 2 \\
Emotet Campaigns                  & 18 & \cmark & 2 & 0 & 2 \\
Silence Trojan                    & 23 & \cmark & 3 & 0 & 5 \\
Elfin / APT33                     & 23 & \cmark & 2 & 0 & 2 \\
SpeakUp                           & 20 & \cmark & 2 & 0 & 1 \\
TSCookie                          & 15 & \cmark & 1 & 0 & 0 \\
Sodinokibi Crescendo			  & 55 & \cmark & 3 & 2 & 6 \\
Sofacy Fysbis                     & 12 & \cmark & 1 & 0 & 0 \\
Gamaredon COVID-19                & 14 & \cmark & 1 & 0 & 2 \\
BabyShark                         & 19 & \cmark & 2 & 0 & 1 \\
Deep Panda			              & 20 & \xmark & 0 & 1 & 0 \\
HermeticWiper                     & 17 & \cmark & 1 & 0 & 1 \\
Rover Trojan                      & 32 & \cmark & 1 & 1 & 1 \\
DEV-0586 / WhisperGate            &  9 & \cmark & 1 & 0 & 2 \\
Sednit Zebrocy                    & 26 & \cmark & 1 & 0 & 4 \\
\midrule
Total                             & 436 & 19/20 & 34 & 4 & 34 \\
\bottomrule
\end{tabular}
\end{table*}

Across the 20 reports, forward chaining derives the specified attack goal in 19 cases, and backward search recovers 34 supporting attack paths in total. 
These results show that the extracted rules usually contain at least one connected path from the initial facts to a security-relevant terminal state. 
The number of paths varies across reports because some reports describe alternative delivery, execution, or payload branches, whereas others contain only one route to the
selected goal. 
The path count is therefore a description of the generated chain structure rather than a higher-is-better metric.

Seventeen of the 20 reports contain no unsupported preconditions in their final chains. 
Only four unsupported preconditions remain across the test: two in Sodinokibi Crescendo, one in Deep Panda, and one in Rover Trojan. 
Sodinokibi Crescendo and Rover Trojan remain goal-reachable because the unresolved conditions are not required by at least one successful path. 
This result shows that a selected goal can remain reachable even when another branch contains an unresolved dependency.

Among the 436 final attack units, 402 units contain at least one postcondition and therefore generate at least one Datalog rule. 
The remaining 34 units have an empty postcondition set and do not produce a rule head. 
These units may preserve behaviors recovered from the report, but they do not define complete state transitions and cannot propagate new facts during forward chaining. 
A NoPost unit may arise because the report does not clearly describe the resulting state,
or because an implicit effect is not successfully mapped to a normalized predicate.

Sodinokibi Crescendo illustrates a different limitation from ordinary dependency breaks. 
The report summarizes several affiliate groups, campaign periods, and initial-access methods, including spear-phishing, RDP compromise, and ransomware distribution through managed service providers. 
These activities are not presented as one continuous attack instance. 
However, the normalization stage fails to preserve campaign-specific entity identities and maps hosts from different incidents to shared symbols such as \texttt{victim\_host}.
This collapses otherwise separate campaign contexts and allows a state from one campaign to incorrectly satisfy a precondition from another.
The resulting cross-campaign paths are not supported by any single attack instance described in the report.
This effect is visible in the three recovered paths. 
For example, one path joins the MSP-distribution branch with an RDP-credential branch
and a local-exploit privilege-escalation branch before reaching the ransomware-deployment goal. 
The individual steps are supported by the report, but the report does not explicitly describe them as one end-to-end attack sequence. 
The resulting path is valid under the generated rules, but it may represent a composition of different campaign fragments rather than one observed attack instance. 
The 2 residual preconditions and 6 NoPost units further show that the report cannot be represented as a single complete chain without preserving campaign-specific context.

Deep Panda is the only report for which the specified goal is unreachable. 
The failure is caused by an implicit static environment condition being represented as a dynamically produced attack state.
The report states that the attackers used WMI to remotely deploy PowerShell scripts and create scheduled tasks on other systems. 
This behavior reasonably requires WMI connectivity between the involved hosts. 
The extracted remote-deployment unit is represented as:

{\scriptsize
\[
\begin{aligned}
\textit{Pre:}\quad
&
\begin{array}[t]{@{}l@{}}
\texttt{access.credential(compromised\_user\_account, password)},\\
\texttt{net.lateral\_access(victim\_host,target\_windows\_host, WMI)}
\end{array}
\\[3pt]
\textit{Post:}\quad
&
\begin{array}[t]{@{}l@{}}
\texttt{access.persistence(target\_windows\_host, scheduled\_task)},\\
\texttt{data.script\_on\_disk(target\_windows\_host,}\\
\qquad\texttt{powershell\_task\_script, PowerShell)}
\end{array}
\end{aligned}
\]
}

The extracted unit uses \texttt{net.lateral\_access} \texttt{(victim\_host, target\_windows\_host, WMI)} as a precondition of the remote-deployment step. 
This predicate is a reasonable prerequisite, but it represents a static target-system condition and should be included in initial fact $F_0$. 
However, in the extracted chain, it is neither included in $F_0$ nor produced by an earlier attack unit.
The diagnosis stage identifies this unsupported precondition, but the repair stage handles it incorrectly. 
It introduces a bridge step that attempts to derive the WMI lateral-access predicate from:

{\scriptsize
\[
\begin{array}{@{}ll@{}}
\textit{Pre:} &
\texttt{net.same\_subnet(victim\_host,target\_windows\_host)}
\\[3pt]
\textit{Post:} &
\texttt{net.lateral\_access(victim\_host,target\_windows\_host, WMI)}
\end{array}
\]
}

This newly introduced precondition is also absent from $F_0$ and is not produced by any preceding unit.
Consequently, the bridge step cannot be executed, and the required WMI lateral-access state is still not derived. 
Forward chaining therefore stops before the remote-deployment step, preventing the subsequent scheduled-task step and the specified goal from being reached.
A more appropriate representation would classify WMI connectivity as a static target-system condition rather than a state produced during the attack. 
Since the report describes the attackers using WMI but does not identify any earlier step that establishes this connectivity, \texttt{net.access(victim\_host, target\_windows\_host, WMI)} should be included in $F_0$ when this interpretation is supported by the report. 
Such a fact should not be introduced merely to make the goal reachable. 
This case highlights that CTI reports often assume existing environmental capabilities without stating them explicitly. 
Distinguishing such static conditions from dynamically produced attack states is essential for reliable attack-path reasoning.

Overall, the results show that the extracted attack units generally form connected rule sets that support attack-goal reachability and attack-path recovery. 
The remaining limitations mainly arise from implicit target-system assumptions, incomplete state transitions, and reports that aggregate multiple campaign contexts into one narrative.

\subsection{Ablation Study}
\label{sec:ablation}

We conduct the ablation study on Cobalt, DeputyDog, Frankenstein, and OceanLotus~\cite{AttacKG2022}. 
These reports cover different attack-chain structures, including multiple delivery paths, converging infection branches, and platform-specific payload delivery. 
All variants are evaluated using the predicate-level metrics defined in Section~\ref{subsec:rq2}.

We compare the full pipeline with four variants.
\textbf{Full} uses all stages of the proposed framework.
\textbf{End-to-End GPT-4.1}  and
\textbf{End-to-End GPT-5.2} directly generate the final normalized attack chain from each input report. 
The two variants use \texttt{gpt-4.1} and \texttt{gpt-5.2}, respectively.
They use the same output schema and vocabularies as the full pipeline but bypass the intermediate stages of behavior extraction, condition extraction, predicate normalization, and diagnosis-guided repair. 
\textbf{w/o Repair} removes the diagnosis-guided repair stage after predicate normalization. 
\textbf{w/o Behavior} removes the behavior vocabulary during attack-skeleton extraction while retaining the remaining stages.

\begin{table*}[t]
\centering
\caption{Ablation results under exact predicate matching.
R, P, and F1 denote predicate recall, precision, and F1,
respectively.}
\label{tab:ablation_predicate}
\setlength{\tabcolsep}{20pt}
\renewcommand{\arraystretch}{1.0}
\begin{tabular}{llccc}
\toprule
Report & Method & R (\%) & P (\%) & F1 (\%) \\
\midrule

\multirow{5}{*}{Cobalt}
 & Full                & \textbf{96.3} & \textbf{95.0} & \textbf{95.7} \\
 & End-to-End GPT-4.1   & 28.0 & 47.5 & 35.3 \\
 & End-to-End GPT-5.2   & 22.0 & 28.9 & 24.9 \\
 & w/o Repair           & 81.7 & 78.0 & 79.8 \\
 & w/o Behavior         & 35.4 & 68.4 & 46.6 \\
\midrule

\multirow{5}{*}{DeputyDog}
 & Full                & \textbf{88.3} & \textbf{100.0} & \textbf{93.8} \\
 & End-to-End GPT-4.1   & 50.0 & 80.0 & 61.5 \\
 & End-to-End GPT-5.2   & 71.7 & 74.5 & 73.0 \\
 & w/o Repair           & 71.7 & 80.8 & 75.9 \\
 & w/o Behavior         & 60.0 & 77.8 & 67.7 \\
\midrule

\multirow{5}{*}{Frankenstein}
 & Full                & \textbf{87.5} & \textbf{100.0} & \textbf{93.3} \\
 & End-to-End GPT-4.1   & 45.3 & 80.0 & 57.9 \\
 & End-to-End GPT-5.2   & 60.9 & 68.4 & 64.5 \\
 & w/o Repair           & 75.0 & 96.4 & 84.4 \\
 & w/o Behavior         & 64.1 & 58.9 & 61.4 \\
\midrule

\multirow{5}{*}{OceanLotus}
 & Full                & \textbf{91.1} & \textbf{100.0} & \textbf{95.3} \\
 & End-to-End GPT-4.1   & 26.7 & 50.0 & 34.8 \\
 & End-to-End GPT-5.2   & 42.2 & 43.2 & 42.7 \\
 & w/o Repair           & 75.6 & 84.2 & 79.6 \\
 & w/o Behavior         & 48.9 & 65.4 & 55.9 \\
\midrule

\multirow{5}{*}{Average}
 & Full                & \textbf{90.8} & \textbf{98.8} & \textbf{94.5} \\
 & End-to-End GPT-4.1   & 37.5 & 64.4 & 47.4 \\
 & End-to-End GPT-5.2   & 49.2 & 53.8 & 51.3 \\
 & w/o Repair           & 76.0 & 84.8 & 79.9 \\
 & w/o Behavior         & 52.1 & 67.6 & 57.9 \\
\bottomrule
\end{tabular}
\end{table*}

Table~\ref{tab:ablation_predicate} shows that the full pipeline
achieves the best result on all four reports. Its average recall,
precision, and F1 are 90.8\%, 98.8\%, and 94.5\%, respectively,
indicating that the pipeline recovers most annotated predicates
while introducing few unsupported predictions.

The two end-to-end variants perform substantially worse.
End-to-End GPT-4.1 achieves an average F1 of 47.4\%. 
End-to-End GPT-5.2 improves recall from 37.5\% to 49.2\%, but its precision decreases from 64.4\% to 53.8\%, resulting in an average F1 of only 51.3\%. 
These results show that using a stronger model alone does not provide the structured and precise predicate extraction achieved by the staged pipeline.

Removing repair reduces the average F1 from 94.5\% to 79.9\%.
The performance decrease appears on all four reports, showing that the repair stage corrects predicate errors that remain after normalization. 
Removing behavior guidance causes a larger decrease, reducing the average F1 to 57.9\%. Without behavior
guidance, the extracted steps have less stable boundaries, which affects subsequent condition extraction and predicate normalization.

Overall, the results show that staged extraction, behavior guidance, and diagnosis-guided repair all contribute to the accuracy of the generated attack chains.

\section{Discussion and Limitations}
\label{sec:discussion}

This work focuses on converting unstructured CTI reports into reachable attack chains. 
By representing each attack step with preconditions, an attack behavior, and postconditions, the framework connects CTI extraction with Datalog-based reachability reasoning.
The evaluation shows that this representation can recover attack behaviors, construct state dependencies, and support attack-path analysis. 
However, several limitations remain.

First, the predicate vocabulary construction still depends heavily on expert knowledge~\cite{mitre2026attackenterprisea} and manual design. 
Different choices of predicate granularity may lead to different representations of the same attack process. 
A coarse vocabulary may merge distinct states, while an overly detailed vocabulary increases the difficulty of normalization and reduces predicate reuse. 
The current vocabulary may therefore require further adjustment when applied to new report types or attack domains.

Second, even with a fixed vocabulary, the LLM does not always map semantically equivalent states to a single canonical representation.
Differences in predicate names, entity aliases, parameter values, and abstraction levels may result in the same preconditions and postconditions being expressed in multiple forms.
These variations reduce exact predicate agreement even when the underlying attack behavior has been correctly recovered.

Third, extraction quality is also limited by the reasoning capability of the
LLM. 
Complex reports often contain implicit conditions, long-range dependencies, and alternative branches. 
The model may omit necessary conditions, merge independent branches, or introduce bridge states that appear internally consistent but are not fully supported by
the report. 
Such errors may leave the generated chain connected while still introducing incorrect assumptions.

Future work will explore more systematic vocabulary construction, stronger entity and predicate canonicalization, evidence-constrained extraction to reduce unsupported model inference, and privacy-preserving inference for hosted LLM services~\cite{ye2026reconstruction} to limit data exposure in sensitive deployments.

\section{Conclusion}
\label{sec:conclusion}

This paper presents an automated framework for transforming unstructured CTI reports into reachable attack chains. 
It represents each attack step with explicit preconditions, an attack behavior, and
postconditions, and converts the normalized attack units into Datalog-style rules. This representation enables CTI knowledge to move beyond flat entities and technique labels toward connected state transitions that support reachability reasoning and attack-path recovery. 
Experiments on 20 CTI reports show that the framework can recover attack behaviors and construct reasoning-ready attack chains more effectively than representative extraction methods.
Overall, this work provides a practical way to turn natural-language CTI
into structured knowledge that can be directly used for automated
security reasoning.

\section*{Ethical Considerations}
This work uses publicly available CTI reports and focuses on extracting structured attack-chain representations for defensive analysis. The framework does not introduce new exploits, malware implementations, or operational attack procedures. The extracted attack units are normalized into abstract predicates and are used for dependency checking and reachability reasoning rather than for executing attacks. Nevertheless, CTI reports may contain sensitive technical details. We therefore avoid releasing additional offensive instructions beyond what is already publicly available in the source reports. The intended use of the framework is to support defensive reasoning, security analysis, and research reproducibility.



%


\bibliographystyle{IEEEtran}
\bibliography{references}


\appendix

\subsection{Behavior Vocabulary}
\label{app:behavior_vocab}

The behavior vocabulary contains 16 coarse-grained behavior classes used
during attack behavior skeleton extraction. These classes help keep attack
steps at a stable granularity, but they are not used in Datalog reasoning.
Table~\ref{tab:behavior_vocab} lists the behavior classes.

\begin{table*}[t]
\centering
\caption{Behavior vocabulary used for attack behavior skeleton extraction.}
\label{tab:behavior_vocab}
\setlength{\tabcolsep}{4pt}
\small
\begin{tabular}{p{0.27\linewidth} p{0.63\linewidth}}
\toprule
Behavior class & Description \\
\midrule
\texttt{user\_action} & A victim-triggered action that directly advances the attack chain, such as opening a document or clicking a link. \\
\texttt{delivery} & A malicious artifact, lure, or initial stage is delivered into the victim environment. \\
\texttt{download} & A document, script, payload, or other artifact is retrieved from a remote location. \\
\texttt{write} & An artifact is written, dropped, or stored on disk, in the registry, or in another persistent location. \\
\texttt{execute} & Code, script, exploit, macro, payload, or command execution occurs. \\
\texttt{evasion} & A defense, policy, restriction, or visibility mechanism is bypassed, weakened, or evaded. \\
\texttt{communication} & The malware or attacker communicates over the network, especially with C2 infrastructure. \\
\texttt{collection} & The malware or attacker gathers information from the victim environment. \\
\texttt{exfiltration} & Previously collected data is sent out of the victim environment to attacker-controlled infrastructure. \\
\texttt{persistence} & An automatic execution mechanism is established or modified to survive reboot, login, or service restart. \\
\texttt{discovery} & Hosts, accounts, services, drives, or network topology are enumerated to guide later movement or targeting. \\
\texttt{credential\_access} & Credentials, tokens, password hashes, or cookies are obtained, dumped, reused, or brute-forced. \\
\texttt{lateral\_movement} & The attacker moves to another host, account, or network segment via valid credentials or remote execution. \\
\texttt{impact} & Destructive, disruptive, or monetization-oriented effects are caused on the victim environment. \\
\texttt{removable\_media} & Removable media is used as an execution, propagation, command-transfer, or exfiltration channel, especially for air-gapped environments. \\
\texttt{outcome} & A resulting attacker advantage, final compromise state, or achieved end state is explicitly described. \\
\bottomrule
\end{tabular}
\end{table*}

\subsection{Predicate Vocabulary}
\label{app:predict_vocab}

The predicate vocabulary contains 139 predicate templates in 10 categories:
\texttt{net}, \texttt{host}, \texttt{access}, \texttt{data}, \texttt{user},
\texttt{evasion}, \texttt{persistence}, \texttt{discovery}, \texttt{load},
and \texttt{impact}. Each predicate has the form
\texttt{category.predicate(arg1,arg2,...)} and is used to represent a
symbolic precondition or postcondition. Tables~\ref{tab:core_predicate_templates}
and~\ref{tab:aux_predicate_templates} list representative predicate templates.
The complete vocabulary is released with the artifact.

\begin{table*}[t]
\centering
\caption{Core predicate templates for normalized attack units.}
\label{tab:core_predicate_templates}
\setlength{\tabcolsep}{6pt}
\small
\begin{tabular}{p{0.12\textwidth} p{0.36\textwidth} p{0.42\textwidth}}
\toprule
Category & Predicate template & Meaning \\
\midrule
Host state &
\texttt{host.platform(host,platform)} &
The host runs a specific platform or operating system. \\

Host state &
\texttt{host.software(host,software)} &
The host has a required software component. \\

Host state &
\texttt{host.vuln\_present(host,cve)} &
The host is affected by a vulnerability. \\

Host state &
\texttt{host.feature\_enabled(host,feature)} &
A required feature or configuration is enabled. \\

Network state &
\texttt{net.access(src,dst,proto)} &
The source can reach the destination over a protocol. \\

Network state &
\texttt{net.traffic\_allowed(src,dst,proto)} &
Traffic between two endpoints is allowed. \\

User action &
\texttt{user.opens\_file(user,file)} &
The victim opens a file. \\

User action &
\texttt{user.clicks\_link(user,context)} &
The victim clicks a link in an email, document, or web page. \\

Delivery &
\texttt{data.email\_delivered(host,artifact)} &
A malicious email or artifact is delivered to the victim. \\

Document state &
\texttt{data.document\_downloaded(host,doc)} &
A document is downloaded to the host. \\

File state &
\texttt{data.file\_on\_disk(host,file)} &
A file or artifact exists on disk. \\

Execution state &
\texttt{access.process\_started(host,proc)} &
A process is started on the host. \\

Code execution &
\texttt{access.code\_exec(host,priv)} &
The attacker obtains code execution capability. \\
\bottomrule
\end{tabular}
\end{table*}

\begin{table*}[t]
\centering
\caption{Auxiliary and impact predicate templates for normalized attack units.}
\label{tab:aux_predicate_templates}
\small
\begin{tabular}{p{0.13\textwidth} p{0.43\textwidth} p{0.38\textwidth}}
\toprule
Category & Predicate template & Meaning \\
\midrule
Script state &
\texttt{data.script\_on\_disk(host,script)} &
A script is written to or present on disk. \\

Script execution &
\texttt{data.script\_host\_exec(host,proc,type)} &
A script is executed through a host process. \\

Payload state &
\texttt{data.payload\_downloaded(host,payload)} &
A payload is downloaded to the host. \\

Payload state &
\texttt{data.payload\_in\_memory(host,payload)} &
A payload is staged or loaded in memory. \\

Payload execution &
\texttt{data.payload\_executed(host,payload)} &
A payload is executed on the host. \\

Loading &
\texttt{load.dll\_loaded(host,dll)} &
A DLL is loaded by a process or the system. \\

Loading &
\texttt{load.driver\_loaded(host,driver)} &
A driver is loaded on the host. \\

Evasion &
\texttt{evasion.applocker\_bypassed(host)} &
Application control restrictions are bypassed. \\

Evasion &
\texttt{evasion.analysis\_avoided(host)} &
Anti-analysis or sandbox-evasion behavior succeeds. \\

Collection &
\texttt{data.collected(host,data)} &
Information, files, or credentials are collected. \\

Communication &
\texttt{data.c2\_request\_sent(host,c2)} &
The host sends a request to a C2 endpoint. \\

Communication &
\texttt{data.c2\_channel(host,c2,proto)} &
A command-and-control channel is established. \\

Persistence &
\texttt{access.persistence(host,method)} &
A persistence mechanism is installed or enabled. \\

Impact &
\texttt{impact.mbr\_corrupted(host)} &
The master boot record is corrupted. \\

Impact &
\texttt{impact.partition\_corrupted(host,part)} &
A disk partition or file system structure is corrupted. \\

Impact &
\texttt{impact.file\_encryption\_attempted(host,file)} &
The malware attempts to encrypt files. \\
\bottomrule
\end{tabular}
\end{table*}

\subsection{Attack-Unit Annotation Schema}
\label{app:annotation_schema}

Each annotated attack unit contains an identifier, an attack behavior, and
two predicate-group fields: \texttt{pre\_groups} and \texttt{post\_groups}.
A precondition or postcondition may have multiple equivalent or alternative
predicate expressions. These predicates are placed in the same group and are
counted as one semantic condition or effect. Different groups are counted as
separate conditions or effects.
Figure~\ref{fig:annotation_schema} shows a simplified example. Long predicate
arguments are shortened for readability.

\begin{figure}[t]
\centering
\begin{minipage}{0.95\linewidth}
\begin{verbatim}
{
  "id": "G03",
  "name": "Payload is downloaded",
  "pre_groups": [
    ["access.code_exec(host,user)"],
    ["net.access(host,c2,HTTP)"]
  ],
  "post_groups": [
    [
      "data.payload_downloaded(host,payload)",
      "data.file_on_disk(host,payload)"
    ]
  ]
}
\end{verbatim}
\end{minipage}
\caption{Simplified attack-unit annotation schema.}
\label{fig:annotation_schema}
\end{figure}

\subsection{Diagnosis and Repair Details}
\label{app:diagnosis}

This appendix lists the issue types and repair actions used in the
diagnosis-guided repair stage. The repair stage follows the diagnosis
results: it only fixes the flagged issues and keeps the remaining attack
chain unchanged.

Table~\ref{tab:diagnosis_repair} summarizes the issue types used in our
implementation. Connectivity issues indicate broken links between a
precondition and earlier states. Semantic issues indicate predicates that
do not match the report evidence or the predicate vocabulary. Branching and
granularity issues describe incorrectly merged, split, or ordered attack
steps. Repair-induced issues are checked in the second diagnosis round.

\begin{table*}[t]
\centering
\caption{Diagnosis issue types and repair actions used in diagnosis-guided repair.}
\label{tab:diagnosis_repair}
\setlength{\tabcolsep}{4pt}
\renewcommand{\arraystretch}{1.05}
\small
\begin{tabular}{p{0.22\textwidth} p{0.18\textwidth} p{0.50\textwidth}}
\toprule
Issue type & Category & Repair action \\
\midrule
Unsupported precondition &
Connectivity &
Add a supported initial fact or connect the precondition to an earlier postcondition. \\

Missing bridge state &
Connectivity &
Add a minimal bridge predicate to an earlier attack unit when supported by the report. \\

Misplaced condition &
Connectivity &
Move the predicate to the attack step where the report evidence supports it. \\

Semantic drift &
Semantic &
Replace the predicate with a faithful template from the predicate vocabulary. \\

Over-specific mapping &
Semantic &
Use a more general predicate when the report does not support the specific state. \\

Missed mapping &
Semantic &
Map an unresolved natural-language condition to an existing predicate template. \\

Overclaimed postcondition &
Semantic &
Remove the unsupported postcondition or replace it with a weaker supported state. \\

Weak postcondition &
Semantic &
Strengthen the postcondition if the report supports a more useful resulting state. \\

Branch logic error &
Branching &
Split alternative branches so that they are not encoded as conjunctive dependencies. \\

Over-compressed step &
Granularity &
Split one coarse step into multiple attack units with separate state transitions. \\

Over-fragmented step &
Granularity &
Merge redundant fragments that do not produce distinct reasoning states. \\

Ineffective repair &
Repair-induced &
Re-check repaired steps and remove repairs that do not resolve the diagnosed issue. \\

Dynamic state as initial fact &
Repair-induced &
Move a dynamic attack state from the initial fact set to the postcondition of an attack step. \\
\bottomrule
\end{tabular}
\end{table*}

We distinguish static requirements from dynamic attack states during repair.
Static requirements, such as software availability, vulnerability existence,
binary availability, and network reachability, can be added to the initial
fact set when they are supported by the report or target-system setting.
Dynamic states, such as downloaded payloads, started processes, established
C2 channels, and collected data, must be produced by earlier attack units.
They are therefore repaired by adjusting earlier postconditions rather than
by adding them to the initial fact set. This prevents attacker-produced
states from being treated as initial system facts.

\end{document}